\documentclass[lettersize,journal]{IEEEtran}
\usepackage{algorithm}
\usepackage{array}
\usepackage[caption=false,font=normalsize,labelfont=sf,textfont=sf]{subfig}
\usepackage{textcomp}
\usepackage{stfloats}
\usepackage{url}
\usepackage{verbatim}
\usepackage{graphicx}
\usepackage{cite}
\usepackage{amsmath,amssymb,amsfonts}
\usepackage{algorithmic}
\usepackage{times}
\usepackage{latexsym}
\usepackage{makecell}

\usepackage{here}
\usepackage{stmaryrd}
\usepackage{multirow}
\usepackage{changepage}
\usepackage{threeparttable}
\usepackage{tabu}
\usepackage{longtable}
\usepackage{float}
\usepackage{colortbl}
\usepackage[dvipsnames,table]{xcolor}
\usepackage{tabularx}

\usepackage{array}
\usepackage[utf8]{inputenc} 
\usepackage[T1]{fontenc}    
\usepackage[hidelinks]{hyperref}       
\usepackage{booktabs}       
\usepackage{nicefrac}       
\usepackage{microtype}      
\usepackage{lipsum}

\begin{document}

\title{A Multimodal Framework for Dementia Detection via Linguistic and Acoustic Representation Learning}

\author{Loukas Ilias, Dimitris Askounis
\thanks{The authors are with the Decision Support Systems Laboratory, School of Electrical and Computer Engineering, National Technical University of Athens, 15780 Athens, Greece (e-mail: lilias@epu.ntua.gr; askous@epu.ntua.gr).}}



\maketitle

\begin{abstract}
Alzheimer’s disease (AD) is a progressive neurodegenerative disorder and the leading cause of dementia, affecting memory, reasoning, communication, and daily functioning. Early diagnosis is particularly important, as timely intervention may help slow cognitive decline and improve patient care. Recent studies have demonstrated that spontaneous speech contains valuable linguistic and acoustic biomarkers associated with dementia. However, existing approaches often rely on independently trained modality-specific models, feature concatenation strategies, ensemble methods, or attention-based fusion mechanisms that do not explicitly maximize the dependency between speech and transcript representations. In this work, we propose a multimodal deep learning framework for automatic dementia detection that jointly exploits speech and transcript information in an end-to-end trainable manner. Specifically, speech recordings are divided into 10-second segments and passed through a pre-trained HuBERT model to extract contextualized acoustic representations. To better capture informative temporal speech characteristics, attentive statistics pooling is employed to aggregate frame-level acoustic embeddings. For the textual modality, transcripts are encoded using a pre-trained BERT model, where the \texttt{[CLS]} token representation is used as the linguistic embedding. The acoustic and textual representations are subsequently combined using an attention-based Audio-Text Fusion (AT-Fusion) mechanism. In addition, we introduce a Mutual Information Neural Estimation (MINE) objective to maximize the mutual information between modalities and improve multimodal representation alignment. The fused multimodal representation is finally used for dementia classification. Experiments conducted on the publicly available ADReSS Challenge and PROCESS-2 dataset demonstrate the effectiveness and robustness of the proposed approach for speech-based dementia assessment.
\end{abstract}


\section{Introduction}
Alzheimer’s disease (AD) is a progressive neurodegenerative disorder and the most common cause of dementia, affecting memory, language, reasoning, and communication abilities. According to the World Health Organization (WHO), more than 55 million people worldwide are currently living with dementia, with the majority residing in low- and middle-income countries \cite{alzheimer_who}. As the global population ages, the prevalence of dementia is expected to increase significantly, posing major healthcare and socioeconomic challenges. One of the early manifestations of AD is the impairment of speech and language abilities. Patients with dementia often experience difficulties in lexical retrieval, semantic organization, sentence construction, and maintaining coherent conversations. In spontaneous speech, these impairments may appear as hesitations, pauses, repetitions, reduced fluency, or the use of vague and semantically empty expressions. Such linguistic and acoustic abnormalities provide valuable biomarkers for automatic dementia detection. Early diagnosis of AD is particularly important, as timely clinical intervention may help slow cognitive decline and improve patient care and quality of life. Traditional diagnostic procedures, however, are often expensive, time-consuming, and dependent on specialized clinical expertise. Consequently, recent research has focused on developing automatic and non-invasive dementia detection systems using speech and language analysis. Recent advances in deep learning and self-supervised representation learning have significantly improved the modeling of both textual and acoustic information.

Existing studies on automatic dementia detection have explored transcript-based, speech-based, and multimodal approaches for identifying cognitive decline from spontaneous speech. Several works rely mainly on linguistic information extracted from transcripts or ASR hypotheses, using transformer-based language models, pause-aware embeddings, or LLM-derived linguistic features \cite{9769980,pan21c_interspeech,10888563,heitz-etal-2025-linguistic}. Although these approaches capture important semantic and lexical abnormalities, they may overlook acoustic biomarkers such as hesitation patterns, prosody, articulation, and voice quality. Other studies focus primarily on acoustic or speech-derived representations, including pre-trained speech embeddings, temporal-context networks, ensemble methods, and acoustic/prosodic/phonological feature integration \cite{gauder21_interspeech,balagopalan21_interspeech,10096253,PRIYANKA2025111078,11358725,10.3389/fnagi.2026.1786269}. However, these methods do not fully exploit the semantic information contained in transcripts. To address this issue, recent multimodal approaches combine acoustic and linguistic representations using attention mechanisms, cross-attention, alignment strategies, LLM-based features, or multi-task learning \cite{10.3389/fnagi.2022.830943,ILIAS2023101485,9926818,ORTIZPEREZ2023126413,10852391,10998944,10878459,10806741,11460500,10.1145/3636534.3649370,botelho25_interspeech}. For example, previous multimodal studies by Ilias and Askounis employed BERT for transcript modeling and AlexNet-based image representations derived from speech spectrograms \cite{10.3389/fnagi.2022.830943,ILIAS2023101485,9926818}. Although these approaches demonstrated strong performance, AlexNet constitutes an older convolutional architecture that has largely been surpassed by modern transformer-based and self-supervised speech representation models in terms of representation quality and modeling capacity. In addition, many existing multimodal systems rely on feature concatenation, independently trained modality-specific encoders, ensemble strategies, or attention-based fusion without explicitly maximizing the statistical dependency between speech and textual representations. Furthermore, ensemble-based and multi-branch systems often increase computational complexity and training time. Therefore, effectively learning compact, aligned, and discriminative multimodal representations remains an open challenge for robust dementia detection.

In order to address the aforementioned limitations, we propose a multimodal deep learning framework for automatic dementia detection that jointly exploits transcript and speech information in an end-to-end trainable manner. First, the speech recordings are divided into fixed-length segments of 10 seconds and passed through the pre-trained HuBERT model for extracting contextualized acoustic representations. To better summarize frame-level speech embeddings and emphasize cognitively informative speech regions, we employ attentive statistics pooling, which computes attention-weighted mean and standard deviation representations. For the textual modality, transcripts are processed using the BERT tokenizer and encoded with a pre-trained BERT model, where the \texttt{[CLS]} token representation is used as linguistic embedding. Next, we introduce an attention-based Audio-Text Fusion (AT-Fusion) mechanism to adaptively combine acoustic and textual representations. In addition, motivated by recent multimodal representation learning approaches, we employ a Mutual Information Neural Estimation (MINE) objective to maximize the mutual information between speech and transcript embeddings, encouraging stronger cross-modal dependency and representation alignment. The fused multimodal representation is subsequently passed through fully connected layers for dementia classification. We evaluate the effectiveness and robustness of the proposed framework on the publicly available ADReSS Challenge and PROCESS-2 datasets. Experimental results demonstrate that the proposed approach effectively captures complementary linguistic and acoustic biomarkers associated with cognitive decline and achieves competitive performance for automatic dementia detection.

Our main contributions can be summarized as follows: 
\begin{itemize}
    \item We employ a pre-trained HuBERT model for speech representation learning together with attentive statistics pooling, enabling the framework to capture informative temporal acoustic characteristics associated with cognitive decline.
    \item We integrate BERT-based transcript representations with acoustic embeddings through an attention-based Audio-Text Fusion mechanism and introduce a Mutual Information Neural Estimation (MINE) objective to maximize the dependency between modalities and improve multimodal representation alignment.
    \item We evaluate the proposed framework on the publicly available ADReSS Challenge and PROCESS-2 dataset, demonstrating the effectiveness of the proposed multimodal approach for dementia detection.
\end{itemize}

\section{Related Work}

\subsection{Linguistic Only}

Recent studies have explored transcript-based and language-oriented approaches for Alzheimer’s disease detection, focusing on linguistic markers extracted from spontaneous speech. These methods are motivated by the fact that cognitive decline is often reflected in language production, including reduced lexical diversity, word-finding difficulties, syntactic simplification, semantic incoherence, and impaired discourse organization. Searle et al. compared several natural language processing techniques using textual transcripts from spontaneous speech for both AD classification and MMSE score prediction \cite{searle20_interspeech}. Their work demonstrated that transcript-based modeling can provide useful diagnostic information, even without direct acoustic features.

Transformer-based models have also been widely investigated for transcript-level dementia detection. Ilias and Askounis proposed an explainable framework for identifying dementia from transcripts using transformer networks, including BERT-based models, multi-task learning, and explainability techniques (i.e., LIME) \cite{9769980}. Their study showed that transformer representations can capture discriminative linguistic patterns associated with dementia, while explainability methods can help identify differences in language use between AD and non-AD participants. Similarly, Pan et al. employed a BERT-based model to extract linguistic information from ASR hypotheses together with confidence scores, achieving strong performance for dementia detection \cite{pan21c_interspeech}.

More recent work has examined how language models can be enriched with additional speech-related information. Pu and Zhang proposed integrating pause information with word embeddings in transformer-based language models for AD detection from spontaneous speech \cite{10888563}. Instead of relying only on lexical content, their method incorporates pause information into the language representation, allowing the model to capture both semantic and temporal aspects of speech.

Large language models have also been used to extract higher-level linguistic features from spontaneous speech transcripts. Heitz et al. employed GPT-4 to derive linguistic features related to known symptoms of Alzheimer’s disease, showing that LLM-based feature extraction can improve AD detection from spontaneous speech \cite{heitz-etal-2025-linguistic}.

\subsection{Acoustics Only}

Recent studies have also investigated speech-based approaches for Alzheimer’s disease detection and cognitive score prediction using acoustic modeling, temporal-context learning, and ensemble methods. Jin et al. proposed CONSEN, a complementary and simultaneous ensemble framework for joint Alzheimer’s disease detection and MMSE score prediction, demonstrating that combining multiple speech-based representations can improve predictive performance \cite{10096253}. Their work highlighted the effectiveness of ensemble learning for jointly modeling diagnostic classification and cognitive assessment tasks.

Several studies have focused on modeling temporal and contextual characteristics of spontaneous speech. Gao et al. introduced a dual-stage time-context network that captures both local and global temporal dependencies in speech recordings using contextual modeling and attention mechanisms \cite{11358725}. Their findings demonstrated that temporal speech dynamics contain valuable indicators of cognitive decline. Similarly, Gauder et al. explored the use of speech embeddings extracted from pre-trained models for Alzheimer’s disease recognition, showing the effectiveness of self-supervised and transfer learning approaches for speech-based dementia detection \cite{gauder21_interspeech}. Balagopalan et al. additionally investigated acoustic-based methods for automatic Alzheimer’s disease detection from spontaneous speech recordings \cite{balagopalan21_interspeech}.

Other recent work has focused on integrating diverse speech-related features for dementia assessment. Priyanka et al. investigated ensemble and deep learning approaches using acoustic features extracted from audio recordings for dementia prediction \cite{PRIYANKA2025111078}. Furthermore, recent studies explored the integration of acoustic, prosodic, and phonological features for automatic Alzheimer’s disease detection, emphasizing the importance of combining complementary speech-derived biomarkers \cite{10.3389/fnagi.2026.1786269}.

\subsection{Acoustic-Linguistic Fusion}

Several multimodal approaches have been proposed to jointly exploit complementary acoustic and linguistic cues for dementia detection. Early multimodal studies investigated the combination of acoustic, lexical, and disfluency-related features using conventional machine learning methods. For example, Balagopalan et al. explored linguistic and acoustic feature representations combined with support vector machine (SVM) classifiers for Alzheimer’s disease detection \cite{balagopalan20_interspeech}. Similarly, Rohanian et al. proposed a multimodal framework integrating acoustic, lexical, and disfluency features, demonstrating the effectiveness of combining heterogeneous speech and language biomarkers for dementia assessment \cite{rohanian20_interspeech}. Another early transformer-based multimodal approach is Dual BERT \cite{10.3389/fcomp.2021.624683}, which combines textual BERT representations with Speech-BERT embeddings based on the Mockingjay model \cite{9054458} for joint speech-text dementia classification.

More recent work has shifted towards transformer-based multimodal architectures and attention mechanisms. Ilias and Askounis investigated multimodal deep learning methods for dementia detection by jointly modeling speech and transcript-based information through attention-based fusions \cite{10.3389/fnagi.2022.830943, ILIAS2023101485}. Their subsequent work further explored transformer-based multimodal architectures and tensor fusion layers for modeling interactions between acoustic and linguistic modalities \cite{9926818}. Ortiz-Perez et al. also proposed a multimodal deep learning framework for predicting signs of dementia from speech and textual information \cite{ORTIZPEREZ2023126413}.

Advanced multimodal fusion techniques have also been investigated in recent years. Pan et al. proposed TSAC-ATT, a two-step attention-based feature combination and cross-attention framework that integrates BERT-based linguistic representations with Wav2vec2.0 acoustic embeddings for speech-based dementia detection \cite{10852391}. Their approach highlighted the importance of explicitly modeling cross-modal interactions between speech and transcript representations through attention-based fusion. Similarly, Zhang et al. introduced DEMENTIA, a hybrid attention-based multimodal and multi-task learning framework that incorporates expert knowledge for Alzheimer’s disease assessment from speech \cite{10806741}. Chang et al. additionally proposed a dual-modal fusion framework for mild cognitive impairment detection based on autobiographical memory, further demonstrating the value of combining acoustic and linguistic modalities for cognitive assessment \cite{10878459}.

Other recent studies investigated the integration of large language models and alignment-based fusion strategies. Casu et al. combined fine-tuned large language models with acoustic features for enhanced AD detection, demonstrating that semantic representations extracted from modern language models can significantly improve multimodal dementia assessment \cite{10998944}. Pan et al. further proposed an unsupervised alignment feature fusion framework for spoken language-based dementia detection, aiming to improve the alignment between acoustic and linguistic representations \cite{11460500}.

Additional multimodal studies have explored transformer-based architectures within federated learning frameworks, as well as ensemble learning strategies, for cognitive impairment detection from spontaneous speech \cite{10.1145/3636534.3649370,botelho25_interspeech}. These works consistently demonstrate that multimodal learning approaches outperform unimodal systems, as speech and language contain complementary dementia-related information.

\subsection{Related Work Review Findings}

Despite these advances, effectively modeling the dependency between acoustic and textual representations remains challenging. Many existing methods rely on feature concatenation, attention-based fusion, or independently learned modality-specific embeddings without explicitly maximizing the shared information between modalities. In contrast, our proposed framework combines BERT-based transcript representations with HuBERT-based acoustic representations, applies attentive statistics pooling to capture informative temporal speech characteristics, and introduces a mutual information maximization objective to encourage stronger cross-modal dependency between speech and language embeddings.

\section{Datasets}

\noindent \textbf{ADReSS Challenge} \cite{luz20_interspeech}. The dataset contains speech recordings and corresponding transcripts from 156 participants, including 78 patients diagnosed with Alzheimer’s disease (AD) and 78 cognitively healthy control subjects.

In the data collection procedure, each participant (PAR) was asked by an interviewer (INV) to describe the Cookie Theft picture from the Boston Diagnostic Aphasia Examination (BDAE) \cite{10.1001/archneur.1994.00540180063015}. The recordings therefore capture spontaneous speech and language patterns commonly used for dementia assessment.

The ADReSS dataset was carefully designed to reduce common sources of bias frequently present in dementia detection studies. In particular, the dataset is balanced with respect to gender and age distribution, and it avoids repeated speech samples from the same participant, which are often encountered in longitudinal datasets. Additionally, the audio recordings were preprocessed to minimize variability in recording quality across subjects.

The dataset is divided into predefined training and test sets. The training set contains 108 participants, including 54 AD patients and 54 healthy controls, while the test set consists of 48 participants, equally divided into 24 AD and 24 non-AD subjects.

\noindent \textbf{PROCESS-2} \cite{pahar2026PROCESS2}. The PROCESS-2 dataset is additionally used to further evaluate the proposed framework on early cognitive impairment detection. PROCESS-2 is a large-scale benchmark speech corpus collected through the CognoMemory digital assessment platform and designed to support automatic cognitive assessment from spontaneous and task-oriented speech. The dataset includes recordings from 400 participants, comprising 200 healthy controls (HC), 150 individuals with mild cognitive impairment (MCI), and 50 individuals diagnosed with Alzheimer’s disease (AD). Each participant completed a single assessment session involving picture description and verbal fluency tasks. The dataset contains approximately 21 hours of speech audio, manually verified transcripts, participant-level metadata, and predefined train/test partitions. Compared with smaller benchmark datasets, PROCESS-2 provides a broader clinical setting for evaluating model robustness across healthy control, mild cognitive impairment, and dementia groups.

In this work, we formulate the PROCESS-2 task as a two-way classification problem, where participants with MCI and AD are grouped into a single cognitively impaired category and compared against healthy controls (HC). We use the Cookie Theft Description task. This setting enables the evaluation of the proposed framework for the detection of early cognitive impairment using multimodal speech and language representations.

The dataset is publicly available on HuggingFace\footnote{https://huggingface.co/datasets/CognoSpeak/PROCESS-2}.

\section{Methodology}

\begin{figure*}
    \centering
    \includegraphics[width=0.95\linewidth]{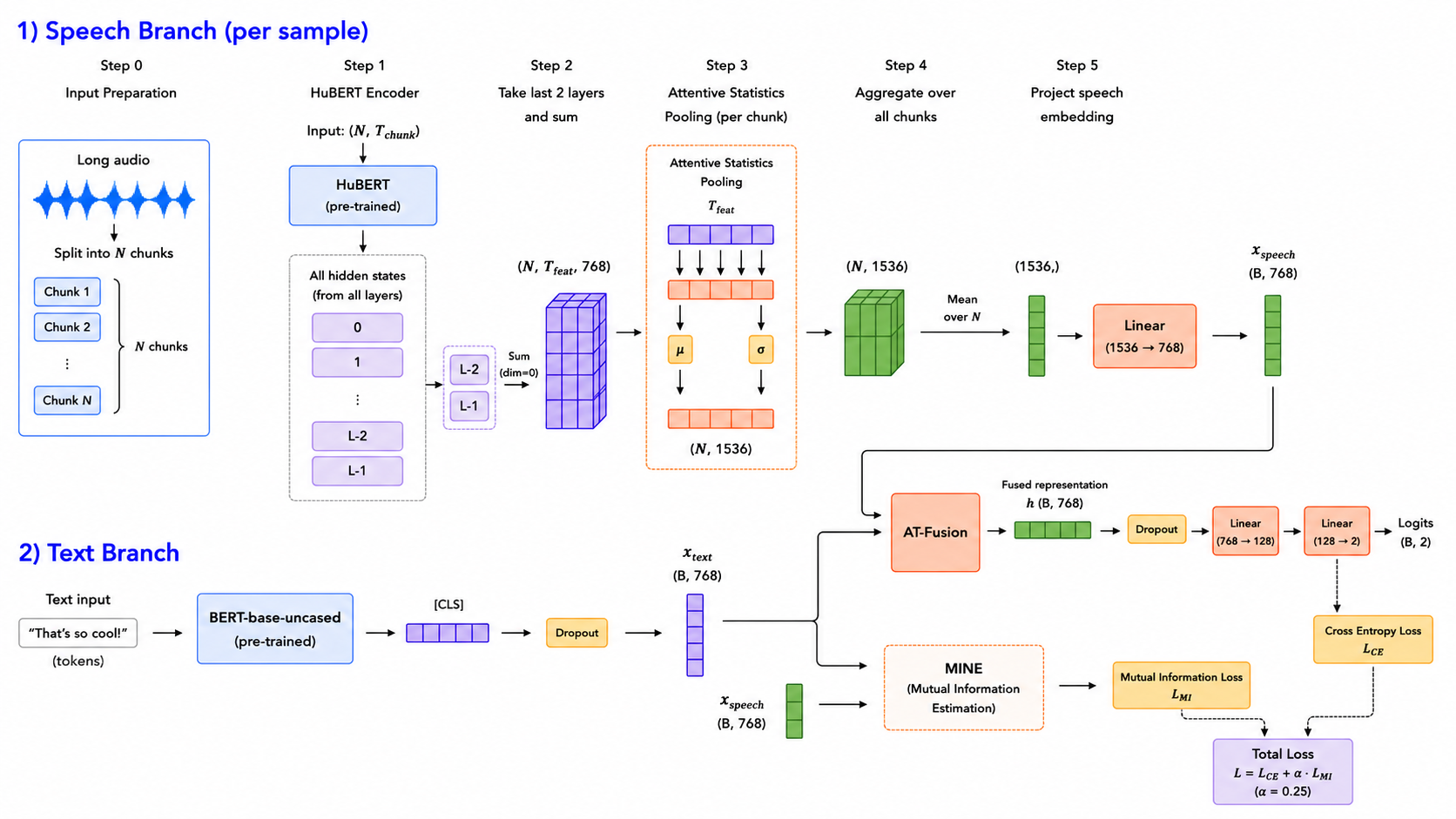}
    \caption{Proposed Methodology}
    \label{methodology}
\end{figure*}

An overview of the proposed multimodal dementia detection framework is presented in Fig.~\ref{methodology}. The individual components and architectural layers of the proposed framework are described in detail below.

\paragraph{Transcript} To extract linguistic representations from speech transcripts, we employ a pre-trained BERT model. The input transcript is first processed using the BERT tokenizer, which converts the text into subword token sequences compatible with the BERT vocabulary. The tokenized sequence is then passed to the BERT encoder to obtain contextualized representations.

Let the transcript be represented as a sequence of tokens
$\{w_1,w_2,\dots,w_N\}$,
where $N$ denotes the number of tokens. BERT produces hidden representations for all tokens:

\[
\mathbf{H}
=
\text{BERT}(w_1,w_2,\dots,w_N)
\]

where
$\mathbf{H} \in \mathbb{R}^{N \times d}$
and $d$ denotes the hidden dimension.

Following standard text classification settings, we use the hidden representation corresponding to the special \texttt{[CLS]} token as the textual embedding:

\[
\mathbf{f}_t
=
\mathbf{H}_{\texttt{[CLS]}}
\]

The \texttt{[CLS]} representation captures the global semantic and contextual information of the transcript and is subsequently used for multimodal fusion with the acoustic representation.

\paragraph{Speech} To extract acoustic representations from raw speech, we employ the pre-trained HuBERT model. The input audio recordings are divided into fixed-length segments of 10 seconds to ensure consistent processing and efficient batch training.

Given an input speech segment
$\mathbf{x} = \{x_1,x_2,\dots,x_T\}$,
the HuBERT encoder produces a sequence of contextualized frame-level acoustic representations:

\[
\mathbf{H}_a
=
\text{HuBERT}(\mathbf{x})
\]

where
$\mathbf{H}_a \in \mathbb{R}^{L \times d}$
denotes the sequence of acoustic embeddings,
$L$ is the number of frame-level representations,
and $d$ is the hidden feature dimension.

Instead of using only the final HuBERT hidden layer, we combine the last two hidden representations to obtain richer contextual acoustic embeddings:

\[
\mathbf{H}_a
=
\mathbf{H}^{(-1)}
+
\mathbf{H}^{(-2)}
\]

where
$\mathbf{H}^{(-1)}$
and
$\mathbf{H}^{(-2)}$
denote the last and second-to-last hidden layers of HuBERT, respectively.

The extracted HuBERT representations capture high-level acoustic and paralinguistic characteristics such as speech fluency, pauses, articulation patterns, and prosodic variations, which are relevant for dementia detection. The frame-level embeddings of each speech segment are subsequently aggregated using attentive statistics pooling to obtain chunk-level acoustic representations.

\noindent \textbf{Attentive Statistics Pooling \cite{okabe18_interspeech}.}
The speech encoder produces a sequence of frame-level acoustic representations
$\{\mathbf{h}_1,\mathbf{h}_2,\dots,\mathbf{h}_T\}$,
where $T$ denotes the number of frames and
$\mathbf{h}_t \in \mathbb{R}^{D}$ represents the feature vector at time step $t$.

Conventional statistics pooling aggregates frame-level features by computing the mean and standard deviation across all frames with equal importance. However, not all speech regions contribute equally to dementia detection. Cognitive impairments are often reflected in specific regions containing hesitations, pauses, articulation abnormalities, or prosodic variations. Therefore, we employ attentive statistics pooling (ASP) to emphasize cognitively informative frames.

An attention mechanism is first applied to estimate the importance of each frame:

\[
e_t
=
\mathbf{v}^{T}
f(\mathbf{W}\mathbf{h}_t+\mathbf{b}) + k
\]

where $\mathbf{W}$ and $\mathbf{b}$ are learnable parameters,
$\mathbf{v}$ is the attention vector,
$k$ is a bias term,
and $f(\cdot)$ denotes a nonlinear activation function.

The attention weights are normalized using a softmax function:

\[
\alpha_t
=
\frac{\exp(e_t)}
{\sum_{\tau=1}^{T}\exp(e_{\tau})}
\]

The weighted mean vector is then computed as:

\[
\tilde{\boldsymbol{\mu}}
=
\sum_{t=1}^{T}
\alpha_t \mathbf{h}_t
\]

In addition to the weighted mean, ASP also computes the weighted standard deviation to capture long-term temporal variability:

\[
\tilde{\boldsymbol{\sigma}}
=
\sqrt{
\sum_{t=1}^{T}
\alpha_t
\mathbf{h}_t \odot \mathbf{h}_t
-
\tilde{\boldsymbol{\mu}}
\odot
\tilde{\boldsymbol{\mu}}
}
\]

where $\odot$ denotes element-wise multiplication.

The final chunk-level acoustic representation is obtained by concatenating the weighted mean and weighted standard deviation:

\[
\mathbf{f}_a
=
[\tilde{\boldsymbol{\mu}},
\tilde{\boldsymbol{\sigma}}]
\]

Since each recording is divided into multiple 10-second segments, the final utterance-level speech representation is computed by averaging the chunk-level embeddings across all segments:

\[
\mathbf{z}_a
=
\frac{1}{N}
\sum_{i=1}^{N}
\mathbf{f}_a^{(i)}
\]

where
$N$
denotes the number of speech segments in the recording.

Compared with average pooling, attentive statistics pooling captures both the dominant acoustic characteristics and the temporal variability of speech signals. This is particularly useful for dementia detection, where acoustic biomarkers may appear intermittently across an utterance. By assigning larger weights to informative speech regions, ASP enables the model to learn more discriminative dementia-related acoustic representations.

\paragraph{MINE \cite{pmlr-v80-belghazi18a}} The acoustic and textual modalities contain complementary dementia-related information. During multimodal fusion, it is desirable for the learned representations from both modalities to preserve correlated semantic and cognitive characteristics. Therefore, the mutual information between the acoustic and textual features should be maximized to improve representation consistency and discriminability. This method has been used successfully in multiple tasks, including multimodal emotion recognition \cite{cai24b_interspeech}.

The mutual information between two random variables $X$ and $Z$ is defined as:

\[
I(X,Z)=\sum_{x \in X}\sum_{z \in Z} p(x,z)
\log \frac{p(x,z)}{p(x)p(z)}
\]

where $p(x,z)$ denotes the joint distribution and $p(x)$ and $p(z)$ are the marginal distributions.

In our framework, $X$ and $Z$ correspond to the acoustic and textual representations extracted from the speech and transcript encoders, respectively. Specifically, let $\mathbf{f}_a$ denote the acoustic embedding extracted from HuBERT and $\mathbf{f}_t$ denote the textual embedding extracted from BERT after attentive pooling.

Direct estimation of mutual information in high-dimensional continuous spaces is intractable because the underlying distributions are unknown. Therefore, we employ Mutual Information Neural Estimation (MINE), which estimates a lower bound of mutual information using a neural network parameterized by $\theta$.

Following the Kullback-Leibler divergence formulation, mutual information can be expressed as:

\[
I(X,Z)=D_{KL}(p(x,z)\|p(x)p(z))
\]

Based on the Donsker-Varadhan representation, the lower bound of mutual information can be written as:

\[
I(X,Z)\geq
\sup_{T \in \mathcal{F}}
\mathbb{E}_{p(x,z)}[T]
-
\log
\left(
\mathbb{E}_{p(x)p(z)}[e^T]
\right)
\]

where $T$ is a statistics network consisting of simple linear layers and ReLU activation functions, which is trained iteratively to estimate and maximize a lower bound of the mutual information between the acoustic and textual modalities. The network learns to distinguish positive acoustic-text pairs sampled from the joint distribution from negative pairs sampled from the product of marginal distributions. If a neural network with parameters $\theta \in \Theta$ is used to fit the function $T$, the Donsker--Varadhan lower bound can be rewritten as:

\[
I(X,Z)
\geq
\sup_{\theta \in \Theta}
\mathbb{E}_{p(x,z)}[T_{\theta}]
-
\log
\left(
\mathbb{E}_{p(x)p(z)}
\left[
e^{T_{\theta}}
\right]
\right)
\]

In the actual training process, the true distributions are unknown and can only be approximated from the sampled data. Let
$\hat{p}_{xz}^{(n)}$,
$\hat{p}_{x}^{(n)}$,
and
$\hat{p}_{z}^{(n)}$
denote the empirical distributions of
$p(x,z)$,
$p(x)$,
and
$p(z)$,
respectively, estimated from $n$ i.i.d. samples. The mutual information lower bound is then estimated as:

\[
\widehat{I(X,Z)}_n
=
\sup_{\theta \in \Theta}
\mathbb{E}_{\hat{p}_{xz}^{(n)}}[T_{\theta}]
-
\log
\left(
\mathbb{E}_{\hat{p}_{x}^{(n)}\hat{p}_{z}^{(n)}}
\left[
e^{T_{\theta}}
\right]
\right)
\]

Replacing $X$ and $Z$ with multimodal representations $\mathbf{f}_a$ and $\mathbf{f}_t$, the multimodal mutual information maximization objective becomes:

\[
\mathcal{L}_{mi} = -\widehat{I(\mathbf{f}_a,\mathbf{f}_t)}_n
\]

\paragraph{Fusion \cite{lian20b_interspeech}} To effectively combine acoustic and linguistic information, we employ an Audio-Text Fusion (AT-Fusion) mechanism to learn modality-aware fusion weights between speech and transcript representations.

Let
$\mathbf{f}_a \in \mathbb{R}^{D}$
denote the acoustic representation obtained from attentive statistics pooling, and
$\mathbf{f}_t \in \mathbb{R}^{D}$
denote the textual representation extracted from the BERT \texttt{[CLS]} token.

The multimodal feature representation is first constructed through concatenation:

\[
\mathbf{f}_{cat}
=
\text{Concat}(\mathbf{f}_a,\mathbf{f}_t)
\in
\mathbb{R}^{D \times 2}
\]

An attention mechanism is then applied to estimate the contribution of each modality:

\[
\alpha_{fuse}
=
\text{softmax}
\left(
\mathbf{w}_f^{T}
\tanh(\mathbf{W}_f \mathbf{f}_{cat})
\right)
\in
\mathbb{R}^{1 \times 2}
\]

where
$\mathbf{W}_f$
and
$\mathbf{w}_f$
are learnable parameters.

Finally, the fused multimodal representation is computed as:

\[
\mathbf{h}
=
\mathbf{f}_{cat}\alpha_{fuse}^{T}
\in
\mathbb{R}^{D \times 1}
\]

The AT-Fusion module enables the model to adaptively emphasize the more informative modality for dementia detection, allowing complementary acoustic and linguistic biomarkers to be jointly represented in the final multimodal embedding.

\paragraph{Classification Layer} The fused multimodal representation
$\mathbf{h}$
is passed to a fully connected dense layer for the final prediction task. For dementia classification, the output layer consists of two output units corresponding to the target classes:

\[
\hat{\mathbf{y}}
=
\text{Dense}(\mathbf{h})
\in
\mathbb{R}^{2}
\]

where
$\hat{\mathbf{y}}$
denotes the predicted class logits.

The final objective function combines the dementia classification loss and the mutual information maximization loss:

\[
\mathcal{L}
=
\mathcal{L}_{cls}
+
\lambda \mathcal{L}_{MI}
\]

where $\mathcal{L}_{cls}$ denotes the cross-entropy classification loss and $\lambda$ is a hyperparameter controlling the contribution of the mutual information regularization term.

By maximizing mutual information between acoustic and linguistic representations, the proposed framework learns more discriminative multimodal embeddings that capture both speech and language biomarkers associated with cognitive decline.

\section{Experiments}

\subsection{Baselines}
\subsubsection{ADReSS Challenge}
We compare our introduced approach with the following research studies:
\begin{itemize}
    \item Dual BERT \cite{10.3389/fcomp.2021.624683}: This method employs a conventional BERT model for textual transcript representations and a Speech-BERT model based on Mockingjay \cite{9054458} for acoustic speech representations extracted from Mel-spectrograms.
    \item TSAC-ATT \cite{10852391}: The method uses BERT to extract text-based representations and Wav2vec2.0 to extract acoustic representations from speech signals. It employs a two-step attention-based feature combination strategy together with cross-attention to model interactions between the two modalities.
    \item Acoustic + Lexical + Dis \cite{rohanian20_interspeech}: This method proposes a multimodal framework that combines acoustic, lexical, and disfluency-related features for Alzheimer’s disease detection. The approach employs separate LSTM networks for the textual and acoustic modalities, while a gating mechanism is used to fuse the unimodal decisions and control the contribution of each modality to the final prediction.
\end{itemize}

\subsubsection{PROCESS-2}
We compare the proposed framework against a BERT-based baseline model. Specifically, each transcript is encoded using the pre-trained BERT model, and the resulting \texttt{[CLS]} token representation is subsequently passed through two fully connected layers consisting of 128 and 2 units, respectively, to produce the final classification output.
\subsection{Experimental Setup}

The ADReSS Challenge training set is divided into training and validation subsets using a 65\%–35\% split. All introduced architectures are trained five times, and the final results on the official test sets are reported as mean $\pm$ standard deviation across the five runs. We use a batch size of 8 for all experiments. Early stopping is applied to avoid overfitting, where training is terminated if the validation loss does not decrease for eight consecutive epochs. In addition, we employ a StepLR learning rate scheduler with a step size of 4 and a decay factor $\gamma$ = 0.1. To improve cross-modal representation learning, we employ a Mutual Information Neural Estimation objective. The contribution of the mutual information loss is controlled by the weighting parameter $\lambda$ and is equal to 0.25. All models are implemented in PyTorch and trained using a single  NVIDIA A100 PCIe 80GB GPU.

\subsection{Evaluation Metrics}
Accuracy, Precision, Recall, Specificity, and F1-score are reported. We report the mean and standard deviation over five runs.

\section{Results}

\noindent \textbf{ADReSS Challenge.} Table \ref{compare3} presents the performance comparison between the proposed multimodal framework and existing state-of-the-art approaches on the ADReSS Challenge test set. The obtained results demonstrate that the proposed approach achieves competitive and robust performance across all evaluation metrics.

More specifically, the proposed framework attains the highest Recall (88.33\%), outperforming all competing approaches. This result indicates that the proposed model is particularly effective at correctly identifying AD patients, which is especially important in clinical screening scenarios where minimizing false negatives is critical. In addition, the proposed approach achieves the best F1-score (84.31\%) and Accuracy (83.33\%) among all compared methods, demonstrating a strong balance between sensitivity and overall classification performance.

Compared with the Dual BERT-based multimodal approach of \cite{10.3389/fcomp.2021.624683}, the proposed framework improves Accuracy from 82.92\% to 83.33\% and increases the F1-score from 81.24\% to 84.31\%. Similarly, compared with the TSAC-ATT framework \cite{10852391}, the proposed method achieves higher Recall, F1-score, and Accuracy. Although the proposed model obtains slightly lower Specificity compared with some competing methods, this behavior is accompanied by substantially improved Recall, indicating that the framework prioritizes the detection of cognitively impaired subjects more effectively.

\begin{table}[!h]
\scriptsize
\centering
\caption{Performance comparison among proposed model and state-of-the-art approaches on the ADReSS Challenge test set. Reported values are mean $\pm$ standard deviation. Results are averaged across five runs.}
\begin{tabular}{lccccc}
\toprule
\multicolumn{1}{l}{}&\multicolumn{5}{c}{\textbf{Evaluation metrics}}\\
\cline{2-6} 
\multicolumn{1}{l}{\textbf{Architecture}}&\textbf{Prec.}&\textbf{Rec.}&\textbf{FS}&\textbf{Acc.}&\textbf{Spec.}\\
\midrule
\multicolumn{6}{>{\columncolor[gray]{.8}}l}{\textbf{Multimodal state-of-the-art approaches (speech and transcripts)}} \\
\textit{Dual BERT \cite{10.3389/fcomp.2021.624683}} & 83.04 & 83.33 & 82.92 & 82.92 & 82.50\\
\hline
\textit{TSAC-ATT \cite{10852391}} & 81.30 & 81.25 & 81.24 & 81.25 & -\\
\hline
\textit{Acoustic + Lexical + Dis \cite{rohanian20_interspeech}} & 81.82 & 75.00 & 78.26 & 79.17 & 83.33\\
\midrule
\multicolumn{6}{>{\columncolor[gray]{.8}}l}{\textbf{Proposed Approach}} \\
 & 79.97 & 88.33 & 84.31 & 83.33 & 76.66 \\
& $\pm$4.40 & $\pm$6.66 & $\pm$1.50 & $\pm$1.32 & $\pm$7.73 \\ 
\bottomrule
\end{tabular}
\label{compare3}
\end{table}

\noindent \textbf{PROCESS-2 Dataset.} Table \ref{Comparison_With_State_of_the_Art_Approaches_process2} presents the performance comparison between the proposed multimodal framework and the transcript-based BERT baseline on the PROCESS-2 dataset. The results demonstrate that the proposed approach achieves competitive and robust performance for the detection of cognitive impairment using spontaneous speech and transcript information.

More specifically, the proposed framework attains the highest Accuracy (81.75\%) and Specificity (83.50\%), outperforming the BERT baseline in both metrics. The improvement in Specificity indicates that the proposed multimodal approach is more effective at correctly identifying healthy control subjects, reducing the number of false positive predictions. In addition, the proposed model achieves higher Precision (82.92\%) compared with the transcript-only BERT baseline, suggesting that the integration of acoustic information contributes to more reliable positive predictions.

Although the BERT baseline achieves slightly higher Recall and F1-score, the differences remain relatively small, indicating that both linguistic and multimodal approaches perform competitively on the PROCESS-2 benchmark. Nevertheless, the improved Accuracy and Specificity of the proposed framework demonstrate the benefit of combining HuBERT-based acoustic representations with BERT transcript embeddings through attention-based multimodal fusion and mutual information maximization.

\begin{table}[!hbt]
\scriptsize
\centering
\caption{Performance comparison among proposed model and state-of-the-art approaches on the PROCESS-2 dataset. Reported values are mean $\pm$ standard deviation. Results are averaged across five runs.}
\begin{tabular}{lccccc}
\toprule
\multicolumn{1}{l}{\textbf{Architecture}}&\textbf{Prec.}&\textbf{Rec.}&\textbf{FS}&\textbf{Acc.}&\textbf{Spec} \\
\midrule
\multicolumn{6}{>{\columncolor[gray]{.8}}l}{\textbf{ Baselines}} \\ 
\textit{BERT} & 80.27 & 83.00 & 81.49 & 81.25 & 79.49 \\ 
& $\pm$1.14 & $\pm$5.10 & $\pm$2.55 & $\pm$2.09 & $\pm$1.87 \\ 
 \midrule
\multicolumn{6}{>{\columncolor[gray]{.8}}l}{\textbf{ Introduced Approach}} \\
  & 82.92 & 80.00 & 81.40 & 81.75 & 83.50 \\
& $\pm$1.76 & $\pm$3.16 & $\pm$2.07 & $\pm$1.87 & $\pm$1.99 \\ 
 \bottomrule
\end{tabular}
\label{Comparison_With_State_of_the_Art_Approaches_process2}
\end{table}

\section{Ablation Study}

In this section, we perform a series of ablation experiments on the ADReSS Challenge dataset.

\subsection{Impact of Pooling}

Table \ref{ablation_pooling} presents the ablation study investigating the impact of different pooling strategies for aggregating frame-level acoustic representations. Specifically, we compare the proposed attentive statistics pooling approach against conventional mean pooling and max pooling methods.

The obtained results demonstrate that the proposed attentive statistics pooling strategy achieves the best overall performance in terms of Recall (88.33\%), F1-score (84.31\%), and Accuracy (83.33\%). In particular, the substantial improvement in Recall compared with both mean pooling and max pooling indicates that attentive statistics pooling is more effective at identifying AD patients. This behavior suggests that the attention mechanism successfully emphasizes cognitively informative speech regions associated with dementia-related acoustic abnormalities.

Although max pooling achieves the highest Precision and Specificity, it exhibits significantly lower Recall, indicating a tendency toward missing cognitively impaired subjects. Mean pooling produces more balanced results but underperforms the proposed methodology across most evaluation metrics. Furthermore, the proposed approach achieves the lowest standard deviation values for Accuracy and F1-score, suggesting improved robustness and stability across multiple runs.

\begin{table}[!htbp]
\scriptsize
\centering
\caption{Ablation Study (pooling). Best results per evaluation metric are in bold.}
\begin{tabular}{lccccc}
\toprule
\multicolumn{1}{l}{}&\multicolumn{5}{c}{\textbf{Evaluation metrics}}\\
\cline{2-6} 
\multicolumn{1}{l}{\textbf{Architecture}}&\textbf{Precision}&\textbf{Recall}&\textbf{F1-score}&\textbf{Accuracy}&\textbf{Specificity}\\
\midrule
\multicolumn{6}{>{\columncolor[gray]{.8}}l}{\textbf{Ablation Experiments}} \\
\textit{Mean Pooling} &82.54 &82.50 &81.51 &81.25 &80.00 \\ 
&$\pm$10.19 &$\pm$8.89 &$\pm$3.45 &$\pm$4.16 &$\pm$13.54 \\ \hline
\textit{Max Pooling} &86.48 &75.83 &80.16 &81.66 &87.50 \\ 
&$\pm$6.16 &$\pm$10.99 &$\pm$7.06 &$\pm$5.17 &$\pm$7.45 \\ 
\midrule
\multicolumn{6}{>{\columncolor[gray]{.8}}l}{\textbf{Proposed Methodology}} \\
\textit{\textbf{Methodology}} & 79.97 & 88.33 & 84.31 & 83.33 & 76.66 \\
& $\pm$4.40 & $\pm$6.66 & $\pm$1.50 & $\pm$1.32 & $\pm$7.73 \\
\bottomrule
\end{tabular}
\label{ablation_pooling}
\end{table}

\subsection{SSL Speech Models}

Table \ref{ablation_SSL} presents the ablation study investigating the impact of different self-supervised learning (SSL) speech representation models within the proposed multimodal framework. Specifically, we compare the proposed HuBERT-based architecture against wav2vec2.0 \cite{10.5555/3495724.3496768} and XLS-R \cite{babu22_interspeech} speech encoders.

The obtained results demonstrate that the proposed HuBERT-based methodology achieves the best overall performance across the majority of evaluation metrics. In particular, the proposed framework attains the highest Recall (88.33\%), F1-score (84.31\%), and Accuracy (83.33\%), indicating that HuBERT representations provide more discriminative acoustic embeddings for dementia detection. The improvement in Recall is especially important in clinical screening scenarios, as it reflects the ability of the model to correctly identify cognitively impaired subjects.

The wav2vec2.0-based approach achieves the highest Precision and Specificity; however, it exhibits substantially lower Recall and Accuracy compared with the proposed methodology. This behavior suggests that wav2vec2.0 tends to be more conservative in predicting AD cases, resulting in a larger number of false negatives. On the other hand, XLS-R achieves competitive Recall performance but underperforms in Precision, Accuracy, and Specificity, indicating reduced robustness for the considered task.

In addition, the proposed HuBERT-based framework demonstrates lower standard deviation values across most evaluation metrics, suggesting improved training stability and generalization capability. Overall, the results indicate that HuBERT provides more effective contextualized acoustic representations for multimodal dementia detection compared with wav2vec2.0 and XLS-R within the proposed architecture.

\begin{table}[!htbp]
\scriptsize
\centering
\caption{Ablation Study (SSL speech). Best results per evaluation metric are in bold.}
\begin{tabular}{lccccc}
\toprule
\multicolumn{1}{l}{}&\multicolumn{5}{c}{\textbf{Evaluation metrics}}\\
\cline{2-6} 
\multicolumn{1}{l}{\textbf{Architecture}}&\textbf{Precision}&\textbf{Recall}&\textbf{F1-score}&\textbf{Accuracy}&\textbf{Specificity}\\
\midrule
\multicolumn{6}{>{\columncolor[gray]{.8}}l}{\textbf{Ablation Experiments}} \\
\textit{wav2vec2.0} & 83.82 & 75.00 & 77.55 & 77.92 & 80.83 \\ 
& $\pm$12.36 & $\pm$10.54 & $\pm$2.16  & $\pm$5.83  & $\pm$21.98  \\ \hline
\textit{XLS-R} & 75.87 & 86.66 & 80.57 & 79.17& 71.66\\ 
& $\pm$6.06  & $\pm$8.08  & $\pm$4.41 & $\pm$4.56 & $\pm$8.89 \\ 
\midrule
\multicolumn{6}{>{\columncolor[gray]{.8}}l}{\textbf{Proposed Methodology}} \\
\textit{\textbf{Methodology}} & 79.97 & 88.33 & 84.31 & 83.33 & 76.66 \\
& $\pm$4.40 & $\pm$6.66 & $\pm$1.50 & $\pm$1.32 & $\pm$7.73 \\
\bottomrule
\end{tabular}
\label{ablation_SSL}
\end{table}

\subsection{Modifying alpha}

Table \ref{ablation_alpha_mine} presents the ablation study investigating the impact of the weighting parameter $\lambda$ controlling the contribution of the mutual information maximization objective. Specifically, we evaluate the proposed framework under different $\lambda$ values to analyze the effect of the MINE-based regularization term on multimodal representation learning.

The obtained results demonstrate that incorporating the mutual information objective substantially improves the overall performance of the proposed framework. When $\lambda=0$, corresponding to the absence of mutual information maximization, the model achieves the lowest Recall (70.83\%), F1-score (75.96\%), and Accuracy (77.92\%). This behavior indicates that without explicitly encouraging cross-modal dependency, the acoustic and textual representations are less effectively aligned, resulting in reduced multimodal discriminability.

As the value of $\lambda$ increases, the performance improves significantly. In particular, $\lambda=0.2$ achieves the best overall performance, obtaining the highest Recall (87.50\%), F1-score (83.41\%), and Accuracy (82.50\%). These results suggest that moderate mutual information regularization effectively enhances the interaction between acoustic and textual embeddings, allowing the model to better capture complementary dementia-related biomarkers.

Although $\lambda=0$ achieves the highest Precision and Specificity, it exhibits substantially lower Recall, indicating a tendency to miss cognitively impaired subjects. On the other hand, excessively increasing the contribution of the mutual information objective ($\lambda=0.3$) leads to a decrease in overall performance, suggesting that overly strong alignment constraints may negatively affect modality-specific representation learning.

Overall, the results demonstrate that the MINE-based mutual information objective plays an important role in improving multimodal representation alignment and dementia detection performance, with $\lambda=0.25$ providing the best trade-off between sensitivity and overall classification accuracy.

\begin{table}[!htbp]
\scriptsize
\centering
\caption{Ablation Study ($\alpha$). Best results per evaluation metric are in bold.}
\begin{tabular}{lccccc}
\toprule
\multicolumn{1}{l}{}&\multicolumn{5}{c}{\textbf{Evaluation metrics}}\\
\cline{2-6} 
\multicolumn{1}{l}{\textbf{$\lambda$}}&\textbf{Precision}&\textbf{Recall}&\textbf{F1-score}&\textbf{Accuracy}&\textbf{Specificity}\\
\midrule
\textit{0} & 82.94 & 70.83 & 75.96 & 77.92& 85.00 \\ 
& $\pm$5.15  & $\pm$9.50  & $\pm$5.26 & $\pm$4.08 & $\pm$5.65 \\ \hline
\textit{0.1} & 80.75 & 75.83 & 77.93 & 78.75& 81.66 \\ 
& $\pm$2.50  & $\pm$7.64  & $\pm$3.63 & $\pm$2.43 & $\pm$4.25 \\ \hline
\textit{0.2} & 80.25 & 87.50 & 83.41 & 82.50 & 77.50\\ 
& $\pm$6.36 & $\pm$5.89 & $\pm$3.78 & $\pm$4.49 & $\pm$10.41 \\ \hline
\textbf{\textit{0.25}} & 79.97 & 88.33 & 84.31 & 83.33 & 76.66 \\
& $\pm$4.40 & $\pm$6.66 & $\pm$1.50 & $\pm$1.32 & $\pm$7.73 \\ \hline

\textit{0.3} & 78.65 & 79.16& 78.33& 78.33& 77.50\\ 
& $\pm$5.21  & $\pm$9.13  & $\pm$3.00 & $\pm$2.12 & $\pm$8.58 \\ 
\bottomrule
\end{tabular}
\label{ablation_alpha_mine}
\end{table}

\subsection{Fusion Methods}

Table \ref{ablation_fusion} presents the ablation study investigating the impact of different multimodal fusion strategies within the proposed framework. Specifically, we compare the proposed AT-Fusion mechanism against several widely used multimodal fusion approaches, including:

\begin{itemize}
    \item \textbf{Concatenation}
    \item \textbf{Gated Multimodal Unit (GMU) \cite{arevalo2020gated}}: The equations governing the GMU are described as follows: $
    h^t = \tanh{(W^t \hat{E_t}'' + b^t)}, h^v = \tanh{(W^v \hat{E_i}'' + b^v)}, z = \sigma(W^z [\hat{E_t}'';\hat{E_i}''] + b^z), h = z * h^t + (1-z)*h^v$
, where $W^t, W^v, W^z \in \mathbb{R}^{128}$ denote the learnable parameters, and [.;.] the concatenation operation. $h$ is the output of the GMU.
    \item \textbf{MUTAN decomposition \cite{8237547}}: MUTAN is a multimodal fusion method based on bilinear interactions between two modalities. It employs tensor decomposition to efficiently model complex cross-modal relationships while reducing the number of parameters.
    \item \textbf{Multimodal Factorized Bilinear (MFB) pooling \cite{8334194}}: MFB is a bilinear fusion approach that captures multiplicative interactions between multimodal representations through factorized pooling operations. It aims to improve multimodal feature expressiveness while maintaining computational efficiency.
    \item \textbf{Multimodal Factorized High-Order Pooling (MFH) \cite{8334194}}: This is a multimodal fusion method that extends bilinear pooling to capture higher-order interactions between different modalities. The approach stacks multiple factorized bilinear pooling blocks to learn richer and more expressive cross-modal feature representations while maintaining computational efficiency. 
    \item \textbf{BLOCK fusion \cite{Ben-younes_Cadene_Thome_Cord_2019}}: This method is based on the block-term tensor decomposition \cite{doi:10.1137/070690729} and combines the strengths of the Candecomp/PARAFAC (CP) \cite{carroll1970analysis} and Tucker decompositions. 
\end{itemize}

The obtained results demonstrate that the proposed fusion methodology achieves the best overall performance in terms of Recall (88.33\%), F1-score (84.31\%), and Accuracy (83.33\%). In particular, the substantial improvement in Recall compared with all competing fusion methods indicates that the proposed AT-Fusion mechanism is more effective at correctly identifying cognitively impaired subjects.

Although concatenation and GMU \cite{arevalo2020gated} achieve higher Precision and Specificity values, they exhibit considerably lower Recall scores, indicating a tendency to misclassify AD subjects as healthy controls. Similarly, advanced bilinear fusion approaches such as MUTAN \cite{8237547}, MFB, MFH \cite{8334194}, and BLOCK \cite{Ben-younes_Cadene_Thome_Cord_2019} achieve competitive performance but still underperform the proposed methodology in terms of F1-score and Accuracy. Bilinear pooling methods usually introduce a larger number of parameters or impose complex factorization constraints, which can increase the risk of overfitting, especially on small clinical datasets such as ADReSS. This is reflected in the larger standard deviations observed for several fusion baselines. In addition, these fusion approaches primarily model feature interactions at a single representation level and do not explicitly encourage statistical dependency or alignment between acoustic and textual embeddings.

\begin{table}[!htbp]
\scriptsize
\centering
\caption{Ablation Study (Fusion). Best results per evaluation metric are in bold.}
\begin{tabular}{lccccc}
\toprule
\multicolumn{1}{l}{}&\multicolumn{5}{c}{\textbf{Evaluation metrics}}\\
\cline{2-6} 
\multicolumn{1}{l}{\textbf{Architecture}}&\textbf{Precision}&\textbf{Recall}&\textbf{F1-score}&\textbf{Accuracy}&\textbf{Specificity}\\
\midrule
\multicolumn{6}{>{\columncolor[gray]{.8}}l}{\textbf{Ablation Experiments}} \\
\textit{Concatenation} &89.76 &74.16 &81.15 &82.92 &91.66 \\ 
&$\pm$5.79 &$\pm$7.64 &$\pm$6.69 &$\pm$5.80 &$\pm$4.56 \\ \hline
\textit{GMU} &91.02 &68.33 &77.39 &80.42 &92.50 \\ 
&$\pm$5.42 &$\pm$8.98 &$\pm$4.54 &$\pm$2.50 &$\pm$4.86 \\ \hline
\textit{MUTAN} & 81.70& 76.66&78.07 &79.16 &81.66 \\ 
&$\pm$6.40 &$\pm$13.07 &$\pm$6.66 &$\pm$4.75 &$\pm$7.73 \\ \hline
\textit{MFB} & 86.69 & 77.49& 80.79& 82.50& 87.50\\ 
& $\pm$6.39& $\pm$15.28& $\pm$9.88& $\pm$7.41& $\pm$6.97\\ \hline
\textit{MFH} & 83.98& 68.33& 75.00& 77.50& 86.66\\ 
& $\pm$2.79 & $\pm$7.73 &$\pm$4.01 &$\pm$2.43 &$\pm$4.08 \\ \hline
\textit{BLOCK} & 84.16 & 73.33 & 77.30 & 79.16 & 85.00 \\ 
& $\pm$6.60  & $\pm$13.59  & $\pm$6.55  & $\pm$4.75  & $\pm$8.58 \\
\midrule
\multicolumn{6}{>{\columncolor[gray]{.8}}l}{\textbf{Proposed Methodology}} \\
\textit{\textbf{Methodology}} & 79.97 & 88.33 & 84.31 & 83.33 & 76.66 \\
& $\pm$4.40 & $\pm$6.66 & $\pm$1.50 & $\pm$1.32 & $\pm$7.73 \\
\bottomrule
\end{tabular}
\label{ablation_fusion}
\end{table}

\subsection{Investigating Layers of HuBERT}

Table \ref{ablation_hubert_layers} presents the ablation study investigating the impact of combining different numbers of final HuBERT hidden layers for acoustic representation extraction. Specifically, when the number of layers is set to 1, only the last HuBERT hidden layer is used. When it is set to 2, the last two hidden layers are summed, while setting it to 3 corresponds to summing the last three hidden layers.

The obtained results demonstrate that combining the last two HuBERT layers achieves the best overall performance in terms of Recall (88.33\%), F1-score (84.31\%), and Accuracy (83.33\%). This suggests that the last two layers provide complementary acoustic and contextual information that is beneficial for dementia detection. In particular, the substantial improvement in Recall indicates that this configuration is more effective at identifying cognitively impaired subjects.

Although using only the last layer or combining the last three layers achieves higher Precision and Specificity, both configurations obtain lower Recall and F1-score. This indicates that using only the final HuBERT layer may not capture sufficient complementary acoustic information, while incorporating too many final layers may introduce redundant or less task-relevant representations.

\begin{table}[!htbp]
\scriptsize
\centering
\caption{Ablation Study (HuBERT Layers). Best results per evaluation metric are in bold.}
\begin{tabular}{lccccc}
\toprule
\multicolumn{1}{l}{}&\multicolumn{5}{c}{\textbf{Evaluation metrics}}\\
\cline{2-6} 
\multicolumn{1}{l}{\textbf{Architecture}}&\textbf{Precision}&\textbf{Recall}&\textbf{F1-score}&\textbf{Accuracy}&\textbf{Specificity}\\
\midrule
\multicolumn{6}{>{\columncolor[gray]{.8}}l}{\textbf{Layers}} \\
\textit{3} & 84.42 & 79.16& 81.05& 81.66&84.16 \\ 
& $\pm$6.50  & $\pm$8.74  & $\pm$2.29 & $\pm$1.56 & $\pm$8.50 \\ \hline
\textbf{\textit{2}} & 79.97 & 88.33 & 84.31 & 83.33 & 76.66 \\
& $\pm$4.40 & $\pm$6.66 & $\pm$1.50 & $\pm$1.32 & $\pm$7.73 \\  \hline
\textit{1} & 84.44 & 81.66& 82.41& 82.92& 84.16\\ 
& $\pm$5.56 & $\pm$10.74  & $\pm$5.56  & $\pm$4.49  & $\pm$7.64  \\ 
\bottomrule
\end{tabular}
\label{ablation_hubert_layers}
\end{table}

\section{Conclusion}

In this work, we propose a multimodal deep learning framework for automatic dementia detection from spontaneous speech and transcripts. The proposed approach combines BERT-based transcript representations with HuBERT-based acoustic embeddings, while attentive statistics pooling is employed to capture informative temporal speech characteristics. In addition, we introduce an attention-based Audio-Text Fusion mechanism together with a Mutual Information Neural Estimation (MINE) objective to improve multimodal representation alignment. Experimental results on the ADReSS and PROCESS-2 datasets demonstrated the effectiveness and robustness of the proposed framework, achieving competitive performance compared with existing state-of-the-art approaches. Specifically, our proposed approach yields Accuracy up to 83.33\% and 81.75\% on the ADReSS and PROCESS-2 datasets respectively. Finally, ablation studies further highlighted the importance of attentive pooling, mutual information maximization, and the proposed fusion strategy for multimodal dementia detection.

In future work, we aim to investigate additional tasks from the PROCESS-2 dataset, explore more advanced multimodal fusion strategies and contrastive learning approaches, and study federated learning settings where different dementia datasets are treated as independent clients.

\bibliographystyle{unsrt}  
\bibliography{references} 

\begin{thebibliography}{10}

\bibitem{alzheimer_who}
{World Health Organization}.
\newblock {\textit{Dementia}}.
\newblock Available online at: \url{https://www.who.int/news-room/fact-sheets/detail/dementia}, {2021}.

\bibitem{9769980}
Loukas Ilias and Dimitris Askounis.
\newblock Explainable identification of dementia from transcripts using transformer networks.
\newblock {\em IEEE Journal of Biomedical and Health Informatics}, 26(8):4153--4164, 2022.

\bibitem{pan21c_interspeech}
Yilin Pan, Bahman Mirheidari, Jennifer~M. Harris, Jennifer~C. Thompson, Matthew Jones, Julie~S. Snowden, Daniel Blackburn, and Heidi Christensen.
\newblock {Using the Outputs of Different Automatic Speech Recognition Paradigms for Acoustic- and BERT-Based Alzheimer’s Dementia Detection Through Spontaneous Speech}.
\newblock In {\em {Interspeech 2021}}, pages 3810--3814, 2021.

\bibitem{10888563}
Yu~Pu and Wei-Qiang Zhang.
\newblock Integrating pause information with word embeddings in language models for alzheimer’s disease detection from spontaneous speech.
\newblock In {\em ICASSP 2025 - 2025 IEEE International Conference on Acoustics, Speech and Signal Processing (ICASSP)}, pages 1--5, 2025.

\bibitem{heitz-etal-2025-linguistic}
Jonathan Heitz, Gerold Schneider, and Nicolas Langer.
\newblock Linguistic features extracted by {GPT}-4 improve {A}lzheimer{'}s disease detection based on spontaneous speech.
\newblock In Owen Rambow, Leo Wanner, Marianna Apidianaki, Hend Al-Khalifa, Barbara~Di Eugenio, and Steven Schockaert, editors, {\em Proceedings of the 31st International Conference on Computational Linguistics}, pages 1850--1864, Abu Dhabi, UAE, January 2025. Association for Computational Linguistics.

\bibitem{gauder21_interspeech}
Lara Gauder, Leonardo Pepino, Luciana Ferrer, and Pablo Riera.
\newblock {Alzheimer Disease Recognition Using Speech-Based Embeddings From Pre-Trained Models}.
\newblock In {\em {Interspeech 2021}}, pages 3795--3799, 2021.

\bibitem{balagopalan21_interspeech}
Aparna Balagopalan and Jekaterina Novikova.
\newblock {Comparing Acoustic-Based Approaches for Alzheimer’s Disease Detection}.
\newblock In {\em {Interspeech 2021}}, pages 3800--3804, 2021.

\bibitem{10096253}
Longbin Jin, Yealim Oh, Hyunseo Kim, Hyuntaek Jung, Hyo~Jin Jon, Jung~Eun Shin, and Eun~Yi Kim.
\newblock Consen: Complementary and simultaneous ensemble for alzheimer’s disease detection and mmse score prediction.
\newblock In {\em ICASSP 2023 - 2023 IEEE International Conference on Acoustics, Speech and Signal Processing (ICASSP)}, pages 1--2, 2023.

\bibitem{PRIYANKA2025111078}
G.~Priyanka and K.~Amshakala.
\newblock Predicting dementia through audio: Ensemble and deep learning approaches using acoustic features.
\newblock {\em Computers in Biology and Medicine}, 197:111078, 2025.

\bibitem{11358725}
Yifan Gao, Long Guo, and Hong Liu.
\newblock A dual-stage time-context network for speech-based alzheimer’s disease detection.
\newblock {\em IEEE Signal Processing Letters}, 33:788--792, 2026.

\bibitem{10.3389/fnagi.2026.1786269}
M.~Zakaria~Kurdi.
\newblock Integrating acoustic, prosodic, and phonological features for automatic alzheimer’s detection.
\newblock {\em Frontiers in Aging Neuroscience}, Volume 18 - 2026, 2026.

\bibitem{10.3389/fnagi.2022.830943}
Loukas Ilias and Dimitris Askounis.
\newblock Multimodal deep learning models for detecting dementia from speech and transcripts.
\newblock {\em Frontiers in Aging Neuroscience}, Volume 14 - 2022, 2022.

\bibitem{ILIAS2023101485}
Loukas Ilias, Dimitris Askounis, and John Psarras.
\newblock Detecting dementia from speech and transcripts using transformers.
\newblock {\em Computer Speech \& Language}, 79:101485, 2023.

\bibitem{9926818}
Loukas Ilias, Dimitris Askounis, and John Psarras.
\newblock A multimodal approach for dementia detection from spontaneous speech with tensor fusion layer.
\newblock In {\em 2022 IEEE-EMBS International Conference on Biomedical and Health Informatics (BHI)}, pages 1--5, 2022.

\bibitem{ORTIZPEREZ2023126413}
David Ortiz-Perez, Pablo Ruiz-Ponce, David Tomás, Jose Garcia-Rodriguez, M.~Flores Vizcaya-Moreno, and Marco Leo.
\newblock A deep learning-based multimodal architecture to predict signs of dementia.
\newblock {\em Neurocomputing}, 548:126413, 2023.

\bibitem{10852391}
Yilin Pan, Bahman Mirheidari, Daniel Blackburn, and Heidi Christensen.
\newblock A two-step attention-based feature combination cross-attention system for speech-based dementia detection.
\newblock {\em IEEE Transactions on Audio, Speech and Language Processing}, 33:896--907, 2025.

\bibitem{10998944}
Filippo Casu, Andrea Lagorio, Pietro Ruiu, Giuseppe~A. Trunfio, and Enrico Grosso.
\newblock Integrating fine-tuned llm with acoustic features for enhanced detection of alzheimer's disease.
\newblock {\em IEEE Journal of Biomedical and Health Informatics}, pages 1--14, 2025.

\bibitem{10878459}
Ho-Ling Chang, Thiri Wai, Yu-Shan Liao, Sheng-Ya Lin, Yu-Ling Chang, and Li-Chen Fu.
\newblock A dual-modal fusion framework for detection of mild cognitive impairment based on autobiographical memory.
\newblock {\em IEEE Journal of Biomedical and Health Informatics}, 29(6):4474--4485, 2025.

\bibitem{10806741}
Zhenglin Zhang, Tengfei Wang, Zian Hu, Li-Zhuang Yang, and Hai Li.
\newblock Dementia: A hybrid attention-based multimodal and multi-task learning framework with expert knowledge for alzheimer's disease assessment from speech.
\newblock {\em IEEE Journal of Biomedical and Health Informatics}, 29(4):2957--2968, 2025.

\bibitem{11460500}
Yilin Pan, Ziteng Gong, Sui Wang, Zhuoran Tian, Tsy Yih, and Lihe Huang.
\newblock An unsupervised alignment feature fusion system for spoken language-based dementia detection.
\newblock In {\em ICASSP 2026 - 2026 IEEE International Conference on Acoustics, Speech and Signal Processing (ICASSP)}, pages 15597--15601, 2026.

\bibitem{10.1145/3636534.3649370}
Xiaomin Ouyang, Xian Shuai, Yang Li, Li~Pan, Xifan Zhang, Heming Fu, Sitong Cheng, Xinyan Wang, Shihua Cao, Jiang Xin, Hazel Mok, Zhenyu Yan, Doris Sau~Fung Yu, Timothy Kwok, and Guoliang Xing.
\newblock Admarker: A multi-modal federated learning system for monitoring digital biomarkers of alzheimer's disease.
\newblock In {\em Proceedings of the 30th Annual International Conference on Mobile Computing and Networking}, ACM MobiCom '24, page 404–419, New York, NY, USA, 2024. Association for Computing Machinery.

\bibitem{botelho25_interspeech}
Catarina Botelho, David Gimeno-Gómez, Francisco Teixeira, John Mendonça, Patrícia Pereira, Diogo A.~P. Nunes, Thomas Rolland, Anna Pompili, Rubén Solera-Ureña, Maria Ponte, David~Martins de~Matos, Carlos-D. Martínez-Hinarejos, Isabel Trancoso, and Alberto Abad.
\newblock {Acoustic and Linguistic Biomarkers for Cognitive Impairment Detection from Speech}.
\newblock In {\em {Interspeech 2025}}, pages 1418--1422, 2025.

\bibitem{searle20_interspeech}
Thomas Searle, Zina Ibrahim, and Richard Dobson.
\newblock {Comparing Natural Language Processing Techniques for Alzheimer’s Dementia Prediction in Spontaneous Speech}.
\newblock In {\em {Interspeech 2020}}, pages 2192--2196, 2020.

\bibitem{balagopalan20_interspeech}
Aparna Balagopalan, Benjamin Eyre, Frank Rudzicz, and Jekaterina Novikova.
\newblock {To BERT or not to BERT: Comparing Speech and Language-Based Approaches for Alzheimer’s Disease Detection}.
\newblock In {\em {Interspeech 2020}}, pages 2167--2171, 2020.

\bibitem{rohanian20_interspeech}
Morteza Rohanian, Julian Hough, and Matthew Purver.
\newblock {Multi-Modal Fusion with Gating Using Audio, Lexical and Disfluency Features for Alzheimer’s Dementia Recognition from Spontaneous Speech}.
\newblock In {\em {Interspeech 2020}}, pages 2187--2191, 2020.

\bibitem{10.3389/fcomp.2021.624683}
Youxiang Zhu, Xiaohui Liang, John~A. Batsis, and Robert~M. Roth.
\newblock Exploring deep transfer learning techniques for alzheimer's dementia detection.
\newblock {\em Frontiers in Computer Science}, Volume 3 - 2021, 2021.

\bibitem{9054458}
Andy~T. Liu, Shu-wen Yang, Po-Han Chi, Po-chun Hsu, and Hung-yi Lee.
\newblock Mockingjay: Unsupervised speech representation learning with deep bidirectional transformer encoders.
\newblock In {\em ICASSP 2020 - 2020 IEEE International Conference on Acoustics, Speech and Signal Processing (ICASSP)}, pages 6419--6423, 2020.

\bibitem{luz20_interspeech}
Saturnino Luz, Fasih Haider, Sofia de~la Fuente, Davida Fromm, and Brian MacWhinney.
\newblock {Alzheimer’s Dementia Recognition Through Spontaneous Speech: The ADReSS Challenge}.
\newblock In {\em {Interspeech 2020}}, pages 2172--2176, 2020.

\bibitem{10.1001/archneur.1994.00540180063015}
James~T. Becker, François Boiler, Oscar~L. Lopez, Judith Saxton, and Karen~L. McGonigle.
\newblock The natural history of alzheimer's disease: Description of study cohort and accuracy of diagnosis.
\newblock {\em Archives of Neurology}, 51(6):585--594, 06 1994.

\bibitem{pahar2026PROCESS2}
Madhurananda Pahar, Caitlin~H. Illingworth, Bahman Mirheidari, Hend Elghazaly, Fritz Peters, Sophie Young, Wing-Zin Leung, Labhpreet Kaur, Daniel Blackburn, and Heidi Christensen.
\newblock {PROCESS}-2: A benchmark speech corpus for early cognitive impairment detection.
\newblock {\em arXiv preprint arXiv:2605.14888}, 2026.

\bibitem{okabe18_interspeech}
Koji Okabe, Takafumi Koshinaka, and Koichi Shinoda.
\newblock {Attentive Statistics Pooling for Deep Speaker Embedding}.
\newblock In {\em {Interspeech 2018}}, pages 2252--2256, 2018.

\bibitem{pmlr-v80-belghazi18a}
Mohamed~Ishmael Belghazi, Aristide Baratin, Sai Rajeshwar, Sherjil Ozair, Yoshua Bengio, Aaron Courville, and Devon Hjelm.
\newblock Mutual information neural estimation.
\newblock In Jennifer Dy and Andreas Krause, editors, {\em Proceedings of the 35th International Conference on Machine Learning}, volume~80 of {\em Proceedings of Machine Learning Research}, pages 531--540. PMLR, 10--15 Jul 2018.

\bibitem{cai24b_interspeech}
Yunrui Cai, Zhiyong Wu, Jia Jia, and Helen Meng.
\newblock {LoRA-MER: Low-Rank Adaptation of Pre-Trained Speech Models for Multimodal Emotion Recognition Using Mutual Information}.
\newblock In {\em {Interspeech 2024}}, pages 4658--4662, 2024.

\bibitem{lian20b_interspeech}
Zheng Lian, Jianhua Tao, Bin Liu, Jian Huang, Zhanlei Yang, and Rongjun Li.
\newblock {Context-Dependent Domain Adversarial Neural Network for Multimodal Emotion Recognition}.
\newblock In {\em {Interspeech 2020}}, pages 394--398, 2020.

\bibitem{10.5555/3495724.3496768}
Alexei Baevski, Henry Zhou, Abdelrahman Mohamed, and Michael Auli.
\newblock wav2vec 2.0: a framework for self-supervised learning of speech representations.
\newblock In {\em Proceedings of the 34th International Conference on Neural Information Processing Systems}, NIPS '20, Red Hook, NY, USA, 2020. Curran Associates Inc.

\bibitem{babu22_interspeech}
Arun Babu, Changhan Wang, Andros Tjandra, Kushal Lakhotia, Qiantong Xu, Naman Goyal, Kritika Singh, Patrick {von Platen}, Yatharth Saraf, Juan Pino, Alexei Baevski, Alexis Conneau, and Michael Auli.
\newblock {XLS-R: Self-supervised Cross-lingual Speech Representation Learning at Scale}.
\newblock In {\em {Interspeech 2022}}, pages 2278--2282, 2022.

\bibitem{arevalo2020gated}
John Arevalo, Thamar Solorio, Manuel Montes-y Gomez, and Fabio~A Gonz{\'a}lez.
\newblock Gated multimodal networks.
\newblock {\em Neural Computing and Applications}, pages 1--20, 2020.

\bibitem{8237547}
Hedi Ben-younes, Remi Cadene, Matthieu Cord, and Nicolas Thome.
\newblock Mutan: Multimodal tucker fusion for visual question answering.
\newblock In {\em ICCV}, pages 2631--2639, 2017.

\bibitem{8334194}
Zhou Yu, Jun Yu, Chenchao Xiang, Jianping Fan, and Dacheng Tao.
\newblock Beyond bilinear: Generalized multimodal factorized high-order pooling for visual question answering.
\newblock {\em IEEE TNNLS}, 2018.

\bibitem{Ben-younes_Cadene_Thome_Cord_2019}
Hedi Ben-younes, Remi Cadene, Nicolas Thome, and Matthieu Cord.
\newblock Block: Bilinear superdiagonal fusion for visual question answering and visual relationship detection.
\newblock {\em AAAI}, (01), 2019.

\bibitem{doi:10.1137/070690729}
Lieven De~Lathauwer.
\newblock Decompositions of a higher-order tensor in block terms—part ii: Definitions and uniqueness.
\newblock {\em SIMAX}, 30(3):1033--1066, 2008.

\bibitem{carroll1970analysis}
J~Douglas Carroll and Jih-Jie Chang.
\newblock Analysis of individual differences in multidimensional scaling via an n-way generalization of “eckart-young” decomposition.
\newblock {\em Psychometrika}, 35(3):283--319, 1970.

\end{thebibliography}

\end{document}